\RequirePackage{fix-cm}
\documentclass[smallextended]{svjour3}       
\smartqed  
\usepackage{url}
\usepackage{graphicx}
\usepackage{caption}
\usepackage{subcaption}
\usepackage{xcolor}
\usepackage{microtype}
\usepackage{multirow}
\usepackage[inline]{enumitem}
\def\bindi {BINDI}
\usepackage{hyperref}
\usepackage{hyphenat}
\emergencystretch=\maxdimen
\def\imageson{}

\usepackage{natbib}

\begin{document}

\title{End-to-end Recurrent Denoising Autoencoder Embeddings for Speaker Identification
}


\author{Esther Rituerto-González         \and
        Carmen Peláez-Moreno 
}


\institute{E. Rituerto-González \at
              Group of Multimedia Processing,  Department of Signal Theory and Communications, University Carlos III of Madrid, Av. Universidad 30, Leganés, 28911 Madrid, Spain\\
              \email{erituert@ing.uc3m.es}           
           \and
           C. Peláez-Moreno \at
              Group of Multimedia Processing,  Department of Signal Theory and Communications, University Carlos III of Madrid, Av. Universidad 30, Leganés, 28911 Madrid, Spain \\
              \email{carmen@tsc.uc3m.es}
}

\date{Received: date / Accepted: date}

\maketitle

\begin{abstract}
Speech 'in-the-wild' is a handicap for speaker recognition systems due to the variability induced by real-life conditions, such as environmental noise and the emotional state of the speaker. Taking advantage of the principles of representation learning, we aim to design a recurrent denoising autoencoder that extracts robust speaker embeddings from noisy spectrograms to perform speaker identification. The end-to-end proposed architecture uses a feedback loop to encode information regarding the speaker into low-dimensional representations extracted by a spectrogram denoising autoencoder. We employ data augmentation techniques by additively corrupting clean speech with real life environmental noise in a database containing real stressed speech. Our study presents that the joint optimization of both the denoiser and speaker identification modules outperforms independent optimization of both components under stress and noise distortions as well as hand-crafted features.

\end{abstract}

\section{Introduction}
\label{sec:introduction}
Speech in real life is commonly noisy and under unconstrained conditions 
that are difficult to predict and \textcolor{black}{aggravate their recognition and understanding}. Speaker Recognition (SR) systems need high performance under these ‘real-world’ conditions. This is extremely difficult to achieve due to both extrinsic and intrinsic variations and is commonly referred to as Speaker Recognition \textit{in-the-wild}. Extrinsic variations encompass background chatter and music, environmental noise, reverberation, channel and microphone effects, etc. On the other hand, intrinsic variations are the inherent factors to the speakers themselves present in speech, such as age, accent, emotion, intonation or speaking rate \citet{stoll}.


Automatic speech recognition (ASR) systems aim to extract the linguistic information from speech in spite of the intrinsic and extrinsic variations \citet{sota_nist18}. However, \textcolor{black}{speaker recognition (SR) takes advantage of} the intrinsic or idiosyncratic variations to find out the uniqueness of each speaker. Besides intra-speaker variability \textcolor{black}{(emotion, health, age),} the speaker identity results from a complex combination of physiological and cultural aspects. Still, the role of emotional speech has not been deeply explored in SR. Although it could be considered an idiosyncratic trait, it poses a challenge due to the distortions it produces on the speech signal\textcolor{black}{. I}t influences the speech spectrum \textcolor{black}{significantly}, having a considerable impact on the features extracted from it and deteriorating the performance of SR systems. 

At the same time, extrinsic variations have been a long standing challenge affecting the basis of all speech technologies. Deep Neural Networks have given rise to substantial improvements due to their ability to deal with real-world, noisy datasets without the need for handcrafted features specifically designed for robustness. One of the most important ingredients to the success of such methods, however, is the availability of large and diverse training datasets.

\subsection{\textit{BINDI, a cyberphysical system to combat gender-based violence}}

\textcolor{black}{Our goal in this paper is to include a robust Speaker Identification module in \bindi, a cyber-physical  system 
designed to combat gender-based violence within the project EMPATIA\footnote{\url{http://portal.uc3m.es/portal/page/portal/inst_estudios_genero/proyectos/UC3M4Safety}}. }
	
\textcolor{black}{Adopting a multimodal approach \cite{hybrid_bindi}, \bindi~employs intelligent signal processing to autonomously detect a violent situation by means of bio-signals collected by smart sensors --including a microphone-- integrated in wearable edge devices interconnected  through the  smartphone  of  the  user.  One  of  the  edge  devices  is currently conceptualized as a smart bracelet that includes three different physiological  sensors:  Blood  Volume  Pulse  (BVP),  Galvanic Skin  Response  (GSR)  and  Skin  Temperature  (SKT)  \cite{celia3,celia4}. The  second  is  designed  as  a  smart  pendant  equipped  with  a microelectromechanical microphone and able to acquire audio and speech from the environment and the user \cite{ourselved_applied_sciences}. This body area network factor form allows to gather data from the desired sources of information to feed the different intelligent decision engines capable of alerting of a violent situation.}	
		
\textcolor{black}{The duration of the batteries of Bindi has been identified as one of the key features to make it practical for gender-based violence victims (GVV). To this end,  a  lightweight affective computing system based on the bio-signals captured by  the  smart  bracelet  acts  as  a  first  trigger. These bio-signals are periodically captured and processed within this edge device that provides a rapid inference schema.}

\textcolor{black}{Note that the response provided by this first physiological layer is  designed  as  a  low-cost  and  low-consumption  trigger.  Due to  the  importance  of  avoiding  missing  any potentially violent  situation, this  trigger  is  oriented  towards  detecting  any  possible  cues of  fear  or  panic-related  emotions  and  therefore  to  achieve  a high recall rate. However, due to its constrained computational capacity and resources, its precision is not as good as it could be  under  ideal  conditions.  }

\textcolor{black}{Then,   once   an   alarm   is   triggered   by   the   bracelet,   the pendant  begins  to  acquire  audio,  which  is  compressed  and sent  via  Bluetooth  Low  Energy  (BLE)  to  the  smartphone  or smartwatch. The purpose of this acquisition is two-fold: first,this audio is encrypted and sent to a secure server to be used as  an  evidence  in  a  potential  trial  \cite{larrabeiti}  and  second,  it  is  used to  disambiguate  and  contextualise  the  previous  bio-signals and therefore to improve the overall detection rate. Since the processing of the audio signal takes place in the smartphone, it  allows  for  more  energy-consuming  methods.  In  particular, we  aim  at  processing  the  speech  contained  in  the  audio  to identify and track the user’s voice \cite{ourselves_iberspeech} and to detect fear and panic.}

\textcolor{black}{Therefore, the speech information complements that carried by the physiological bio-signals. In this paper, our scope is limited to the processing of the audio modality and in particular to the Speaker Identification (SI) task. In the threatening situations we intend to detect, identifying the user and achieving high SI rates is crucial, and it is most likely that the speaker is under an intense negative emotional state, such as panic, fear, anxiety, or its more moderate relative, stress.} 

\subsection{\textit{Contribution}}

In this paper, we address the combined problem of the lack of environmental noise robustness of SR systems and the effects of \textcolor{black}{negative} emotional speech on their performance. Our contribution capitalizes on using \textcolor{black}{robust speaker discriminator oriented embeddings} extracted from a Recurrent Denoising Autoencoder combined with a \textcolor{black}{Shallow Neural Network --a feed-forward neural network-- acting as a back-end classifier for the task of Speaker Identification}, as detailed in Figure \ref{fig:outline_archit}. This end-to-end architecture is designed to work under adverse conditions, both from the point of view of distorted speech due to stressing situations, and environmental noise. 

We choose speech recorded under spontaneous stress conditions due to its real-life nature. Induced, simulated or acted emotions --especially negative ones-- are known to be perceived and automatically detected much more strongly than real emotions. This suggests that actors are prone to overacting, which casts doubt on the reliability of these samples 
\citet{real_induced_emotions}, being a big drawback for devices working in real life conditions such as \bindi. 

Moreover, we augment our database with synthetic noisy signals by additively contaminating the dataset with environmental noise to \textcolor{black}{emulate speech recorded in real conditions}.

We discuss a recurrent denoising autoencoder architecture based on Gated Recurrent Units (GRU), \textcolor{black}{ where the recurrent architecture targets modelling the temporal context of speech utterances}. The encoder network extracts frame level \textcolor{black}{embeddings} from the speech spectrograms and is jointly optimized with a feed forward network whose output layer calculates speaker class posteriors. \textcolor{black}{With the help of the denoising module that attempts to remove environmental noise information, and the SNN that targets recognizing the speaker, all information that is not directly employed for speakers' identification is dismissed from the embeddings.} In particular, the loss function associated with this last dense network is also fed into the denoising autoencoder to guide its efforts towards the SR task, as will be further described in section \ref{sec:methods}. 

Finally, we put forward that these speaker discrimination oriented embeddings are more robust to noise and stress variability than those optimized separately by comparing the effects of automatically extracted embeddings by this two-stage connected architecture against the two modules separately, hand-crafted features previously demonstrated to be suited for this problem and a frequency recurrent alternative obtained by transposing the inputs to the GRU autoencoder. 
%
%
\textcolor{black}{Moreover, we achieve these results with a computationally lightweight multitask end-to-end architecture that extracts emotion- and noise-robust speaker discriminant embeddings, in spite of the few datasets available for this purpose.}


\section{Related Work}
\label{sec:state_of_the_art}
 
There is a wealth of research aiming at coping with the variability of speech signals. 
Data augmentation is a widely applied technique to enlarge databases with such distortion \cite{da1}, for example by 
adding noise or applying transformations --similarly to the ones introduced by transmission channels--. Speech enhancement techniques are also used to improve the overall perceptual quality of speech, specifically intelligibility \citet{se1, aesr, kerlos_asen}. Remarkably, these techniques can be modified towards 
a speaker recognition objective, instead of audio quality \citet{sota_nist18}. \textcolor{black}{Our difference with respect to similar works such as \citet{voiceIDloss} consists of first, the shallow approach of the back-end oriented towards having a fast and real-time running system in a wearable device, seeking for a balance between computational complexity and performance; and second the use of an end-to-end system for extracting embeddings containing only speaker-relevant information together with the identification task.} 
  
Additionally, in order to alleviate the intrinsic variation mismatch and specifically the one caused by emotions, literature reckons several solutions, such as eliciting emotions in speakers in a way that accomplishes similar effects as spontaneous \citet{iemocap} due to the difficulties of recording authentic emotions --both in terms of privacy and labelling--. Likewise, statistical estimations and domain adaptation methods are also used \citet{li_yuan, ma1, shanin_nassif}. \textcolor{black}{This lack of datasets containing real and natural --not acted-- negative emotions in speech, as the ones a user could experience in a violent situation, is indeed a challenge.} In our previous works (\citet{ourselves_iberspeech,ourselved_applied_sciences}) we explored data augmentation techniques for synthetic \textit{stressed} speech by modifying its pitch and speed with satisfactory results. 
\textcolor{black}{In this work we turn to the development of architectures robust to variability by adapting the architecture rather than the data.}

In speech related applications, several flavours of hand-crafted or manually extracted features have been widely employed in literature, 
\citet{HC_feats,HC_feats_2, GFCC_feats}. 
However, these techniques are labour-intensive and time-consuming, and their generalization abilities and robustness against variability are limited. Thus in the last decade, it has been found that automatically learnt feature representations or DNN-based embeddings are --given the availability of enough training data-- usually more efficient than hand-crafted or manually designed features, allowing to develop better and faster predictive models \citet{rep_learning, cyclic_dda}. Most importantly, automatically learnt feature representations are in most cases more flexible and powerful. 
Representation learning consists on yielding abstract and useful features usually from the signal waveform directly or from relatively sophisticated 
low-dimensional representations, by using autoencoders and other deep learning architectures often generalizing better to unseen data \citet{embeddings1, voiceIDloss}.

Due to the sequential nature of speech signals, their temporal context is of great relevance for classification and prediction tasks \citet{schuller_overview}. Besides, the sequential character of its frequency contents carries very relevant information of speech 
\citet{recurrent_freq2}. Recurrent Neural Networks are powerful tools to model sequential data \citet{rnn_speech}, having become the state of the art due to their improved performance and generalization capabilities. However, the availability of larger databases is, again, of paramount importance for training such networks. Unfortunately, this is not the case of real stressing situations in particular, such as the ones we are facing. 

Recently, performing data augmentation with additive and convolutional noise with neural network embeddings (a.k.a. x-vectors) rise as one of the best approaches in SR. In a wide sense, all of the neural embeddings which include some form of global temporal pooling and are trained to identify the speakers in a set of training recordings are unified under the term x-vectors according to \citet{Snyder2017,Snyder2018b}. Variants of x-vector systems are characterized by different encoder architectures; pooling methods and training objectives \citet{sota_nist18} and in this sense all of the embeddings tested in this paper could be consider such. 


The use of models to effectively denoise --or dereverberate-- speech samples maintaining specific speaker information using DNNs is a flourishing field with emerging work nowadays. 
Current research includes two-stage models showing improved speaker intelligibility 
\citet{spk_enh1}, Long-Short Term Memory architectures exploiting speech sequential characteristics \citet{spk_enh2}, unsupervised feature enhancement modules robust to noise unconstrained conditions \citet{spk_enh3}, and specially targeted speech enhancement modules with the joint optimization of speaker identification and feature extraction modules \citet{spk_enh4},\citet{voiceIDloss},\citet{aesr}. %



\section{Methods}
\label{sec:methods}

The proposed architecture is the combination of a Recurrent Denoising Auto-Encoder (RDAE) and a Shallow fully-connected Neural Network (SNN) backend 
in an end-to-end system.

\ifdefined\imageson
\begin{figure*}[h]
	\centering
	\begin{subfigure}[b]{0.8\textwidth}
		\centering
		\includegraphics[trim={210 130 200 200},clip,width=1\linewidth]{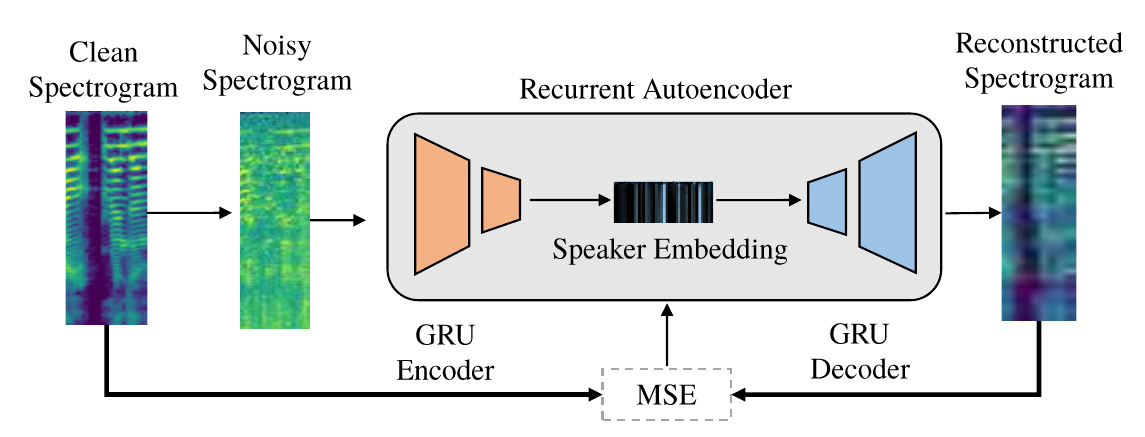}
		\caption{Spectrogram Enhancement}
		\label{fig:sub1}
	\end{subfigure}%
	
	\begin{subfigure}[b]{0.8\textwidth}
		\centering
		\includegraphics[trim={210 150 230 150},clip,width=1\linewidth]{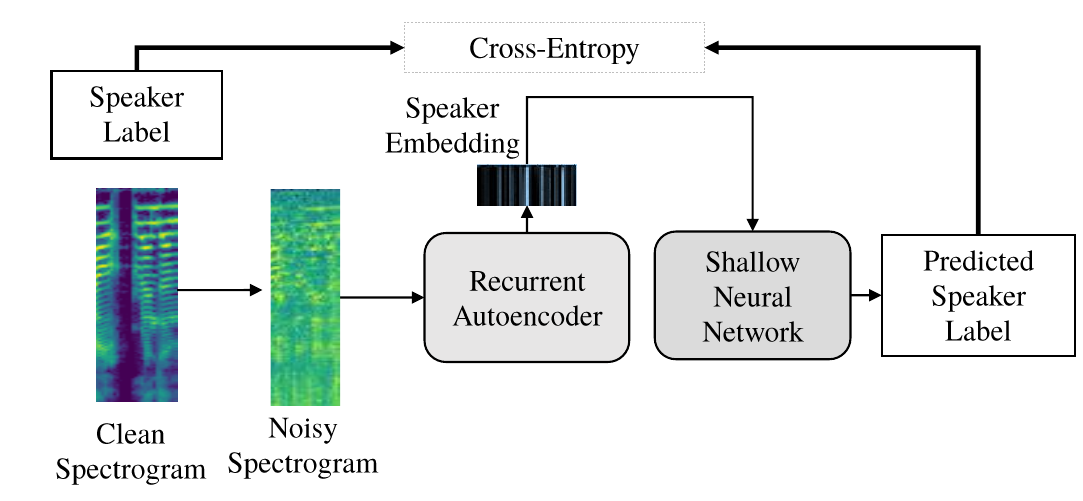}
		\caption{Speaker Identification}
		\label{fig:sub2}
	\end{subfigure}
	\caption{\textcolor{black}{The two components of the Proposed Architecture: the Recurrent Denoising Autoencoder and the SNN}}
	\label{fig:outline_archit}
\end{figure*}
\fi

Autoencoders are generally unsupervised machine learning algorithms trained towards reconstructing their inputs through a series of layers. Denoising Auto-Encoders (DAE) take in a corrupted version of the data as the input and a clean version as the desired output and try to reconstruct the latter from the former. 
Our proposed RDAE is composed of a two-layer encoder and a symmetric decoder based on GRUs. The SNN includes a dense plus a dropout layers, as specified in Figure \ref{fig:outline_archit}.


%

As an input, the encoder takes a one-second log-scaled mel-spectrogram, and encodes it into a low-dimensional representation. 
Although SI systems tend to use longer temporal windows to secure their decisions, \bindi~ needed a real-time and quicker outcome that has motivated this 
\textit{short-utterance} speaker identification architecture. 

After its extraction, the embedding is fed simultaneously to the decoder and the SNN. 
First, the decoder tries to reconstruct a clean spectrogram from this embedding extracted from a noisy spectrum yielding the Mean Squared Error (MSE) 
between the reconstructed and clean spectrograms. Second, the SNN in charge of identifying the speaker to whom that utterance belongs to, computes the cross-entropy of the predicted speaker and the true speaker labels. A block diagram of this architecture can be observed in Figure \ref{fig:outline_archit}. 

Equations \ref{eq:1} and \ref{eq:2} represent the loss functions, $\mathcal L_d$ and $\mathcal L_s$, of the RDAE (mean square error) and SNN (cross-entropy) respectively

\ifdefined\imageson
\begin{equation}
\mathcal L_d = \frac{1}{N} \displaystyle\sum_{i=1}^{N}  (S_{i}-\hat{S}_{i})^{2}
\label{eq:1}
\end{equation}

\begin{equation}
\mathcal L_s = \displaystyle\sum_{i=1}^{N} -logP(\hat{y}_{i} | y_i)
\label{eq:2}
\end{equation}
\fi

\noindent where \textit{S} is the clean spectrogram, $\hat{S}$ the reconstructed spectrogram from the noisy one, and \textit{y} and \textit{$\hat{y}$} are the original and predicted speaker labels. \textit{N} represents the total number of speech samples.
Finally, instead of sequentially training the RDAE and the SNN, the whole architecture is jointly optimized using an equally weighted cost function that linearly combines the previous two metrics as Equation \ref{eq:3}. 

\ifdefined\imageson

\begin{equation}
\mathcal L_T = \lambda \mathcal L_d + (1-\lambda) \mathcal L_s
\label{eq:3}
\end{equation}
\fi

\textcolor{black}{The normalization of the spectrograms results in a normalized MSE loss that falls roughly within the same dynamic range as the cross-entropy loss. Since we did not have any a priori reason to think that one of the tasks could influence the result more that the other we set $\lambda= 0.5$. This showed good results in our test but further exploration of this parameter should be undergone as future work.} 


\section{Experimental Set-up}
\label{sec:experimental_setup}
\subsection{Data}
The VOCE Corpus \citet{voce} is used in this experimentation since first, it contains
data taken in real stress conditions and second, it offers data from sensors similar to those present in \bindi. 
It consists of speech signals from 45 speakers in two different conditions: reading a pre-designed short paragraph and performing an oral presentation presumably causing stress in the speaker. From both settings, voice and heart rate per second are acquired. However, only 21 speakers were finally selected due to incomplete information. 

Each speech signal is labelled with the ID of the speaker every second. The recordings have very different lengths and therefore there is a substantial imbalance in the number of samples per speaker. We decimate the database by choosing approximately 10 minutes of speech per speaker to prevent the model from specializing in majority classes.

The audio recordings from VOCE were converted from stereo to mono and downsampled from 44.1kHz to 16kHz to ease their handling. Also, a normalization fits the signal to the [-1, 1] range. As a final preprocessing step a Voice Activity Detector module (VAD) \citet{voicebox} is applied 
to remove one-second length chunks of non-speech audio where decisions regarding speaker identity cannot be taken.

In order to simulate real-life environments, speech signals were additively contaminated with 5 different noises from -5dB to 20dB in steps of 5dB Signal to Noise Ratios (SNR). Noise signals were chosen from the DEMAND dataset \citet{demand}: 
DWASHING, OHALLWAY, PRESTO, TBUS, SPSQUARE and SCAFE. The noises were chosen to emulate everyday life conditions similar to those envisioned for \bindi~deployment.
The noises were high-pass filtered to eliminate frequencies lower than 60Hz to remove power line interferences, specially noticeable in DWASHING noise. 

We used a 70 ms FFT window, an overlap of 50\% and 140 mel frequency bands and extracted the spectrograms of the speech signals for each second of audio using the spectrogram extraction module in \cite{audeep1,audeep2} thus resulting in 27 time steps and 140 mel-frequency bands mel-spectrograms. These choices showed to be reasonable during a preliminary evaluation. Our choice of a higher number of mel frequency bands and longer temporal windows than typically chosen in hand-crafted feature extraction allows a balance of frequency and time resolution more suited for the recurrent networks. Although the classical choices for these values are inspired in the human auditory system, we hypothesize that machines could take advantage of their computational power when analysing data more than just what humans can hear, and therefore they could be able to overcome the human error rate given enough data is provided.

\subsection{Experiments}
To measure the robustness of the system 
we designed a \textit{multi-conditioning} setting in which all the contaminated speech signals at different SNRs, as well as clean speech signals, are
combined. This is a more realistic scenario in which the specific SNR is not fixed a priori for each training. Special attention was taken to ensure that all samples belonging to the same utterance but contaminated with different noises and SNRs were grouped in the same validation fold, so that none of the various versions of the samples in the validation subset appeared in the training set.

\ifdefined\imageson
\begin{figure}[h]
	\centering
	\begin{subfigure}[b]{1\textwidth}
		\includegraphics[trim=80 270 110 130,clip,width=1\linewidth]{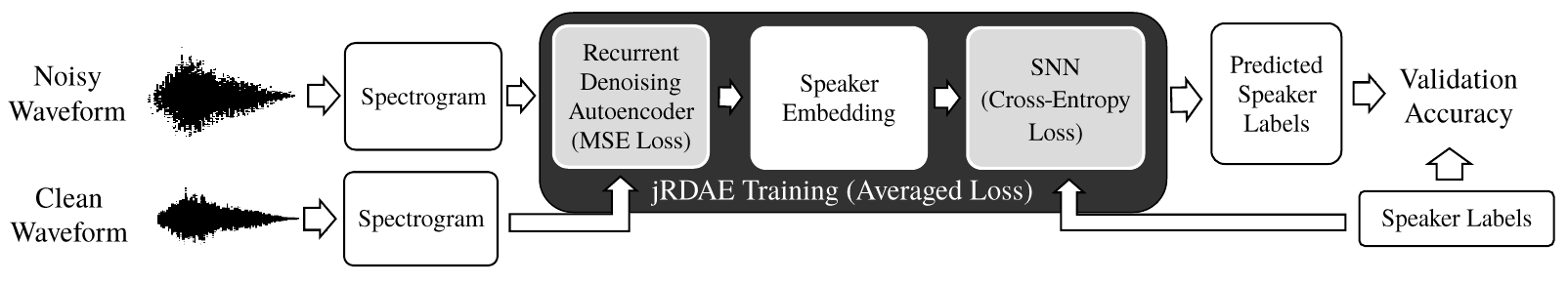}
		\caption{Jointly optimized training procedure}
		
	\end{subfigure}

	\begin{subfigure}[b]{1\textwidth}
		\includegraphics[trim=80 170 110 270, clip,width=1\linewidth]{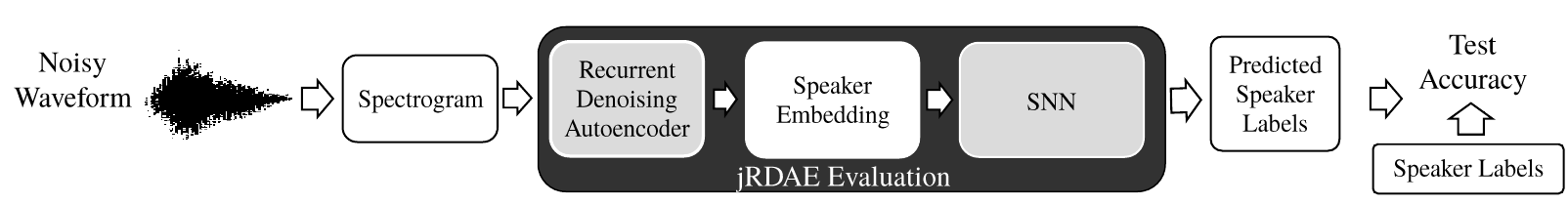}
		\caption{Testing procedure}
		
	\end{subfigure}
	\caption{Training and testing procedures in the proposed architecture.}
	\label{fig:proposed_archit}
\end{figure}
\fi

Nested cross-validation was used to optimize the hyper parameters for the autoencoder and the SNN as speaker classifier. In
nested cross-validation, an outer loop of 33\% of unseen data on the training stage is used to obtain the final test results; an inner loop (3 validation folds) is used to find the optimal hyper parameters via grid search. The test set is unseen so that structural decisions made using data from the same distribution --for which final results are computed-- do not undermine the validity of the conclusions reached. A block diagram of the training and testing procedures can be observed in Figure \ref{fig:proposed_archit}.

The spectrograms are reduced in the frequency axis from $27\times 140$ to $27\times 40$. This low-dimensional image is flattened, obtaining a 1080 one-dimensional speaker embedding. The layer sizes of the architecture are shown in Table \ref{table:autoenc_layers}. 

The number of hidden units of the dense layer of the SNN was set to 1000, dropout percentage to 30\% and the L2 regularization parameter set to 0.01. We trained for 15 epochs with a batch size of 128 and a learning probability of 0.001. We also added a delay to the stop criterion, a patience of 5 iterations, after which if no improvements are observed, training is stopped. 
The model with lower validation loss during the training is selected as the optimal. The spectrograms were normalized with respect to the mean and standard deviation of their training set. 
Each spectrogram in the validation set was normalized in terms of the mean and standard deviation obtained from its correspondent training set in the fold.

\ifdefined\imageson

\begin{table}[h]
	\footnotesize
	\centering
	\begin{tabular}[t]{|c|c|}
		\hline
		\textbf{Layer}&\textbf{Output}\\
		\hline
		Input&(27, 140)\\
		GRU&(27, 64)\\
		GRU&(27, 40)\\
		Flatten&(1080, 1)\\
		\hline
	\end{tabular}
	\hfill
	\begin{tabular}[t]{|c|c|}
		\hline
		\textbf{Layer}&\textbf{Output}\\
		\hline
		Input&(1080, 1)\\
		Reshape&(27, 40)\\
		GRU&(27, 40)\\
		GRU&(27, 64)\\
		\begin{tabular}{@{}c@{}}Time \\ Distributed\end{tabular} &(27, 140)\\
		\hline
	\end{tabular}
	\hfill
	\begin{tabular}[t]{|c|c|}
		\hline
		\textbf{Layer}&\textbf{Output}\\
		\hline
		Input&(1080, 1)\\
		Dense&(1000, 1)\\
		Dropout&(1000, 1)\\
		Dense&(21, 1)\\
		\hline
	\end{tabular}
	\caption{Output dimensions of the layers of the Autoencoder and SNN backend architectures. Encoder (left), decoder (center) and SNN (right).}
	\label{table:autoenc_layers}
\end{table}
\fi


We compared the performance of our proposed \textit{jointly} optimized method (jRDAE) against three different architectures.
First, the same system as ours in which the RDAE and the back-end SNN have been \textit{independently} optimized (iRDAE).
Second, a transposed (frequency) Recurrent Denoising Autoencoder that differs from our approach in that the spectrograms used as input are \textit{transposed}, as well as the GRU layers, and it is the time axis the one reduced in dimensionality. This aims at recurrently modelling the frequency domain. 
Finally, a system in which handcrafted features such as pitch, formants, MFCCs and energy, chosen based in the literature \citet{ourselved_applied_sciences}, are fed directly into the backend SI component, the only module to be trained.



\section{Results \& Discussion}
\label{sec:discussion}
\ifdefined\imageson
\begin{figure*}[!h]
	\centering
	\begin{minipage}[b]{.48\textwidth}
		\vspace{0pt} 
		\includegraphics[width=\columnwidth]
		{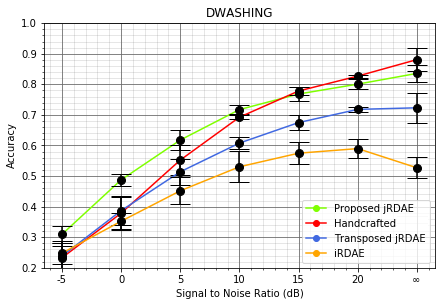}
		\label{label-1}
	\end{minipage}
	\begin{minipage}[b]{.48\textwidth}
		\vspace{0pt} 
		\includegraphics[width=\columnwidth]
		{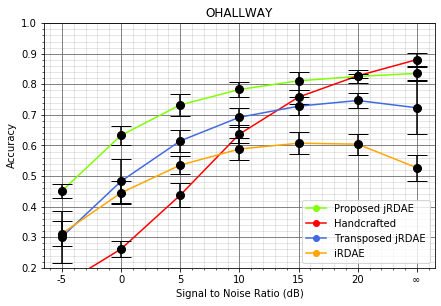}
		\label{label-2}
	\end{minipage}
	\begin{minipage}[b]{.48\textwidth}
		\vspace{0pt}
		\includegraphics[width=\columnwidth]
		{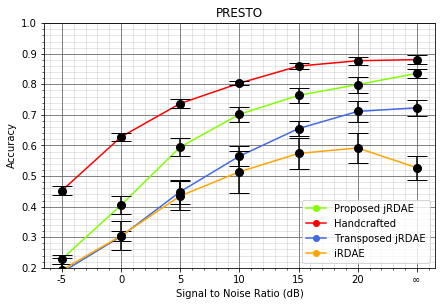}
		\label{label-3}
	\end{minipage}
	\begin{minipage}[b]{.48\textwidth}
		\vspace{0pt}
		\includegraphics[width=\columnwidth]
		{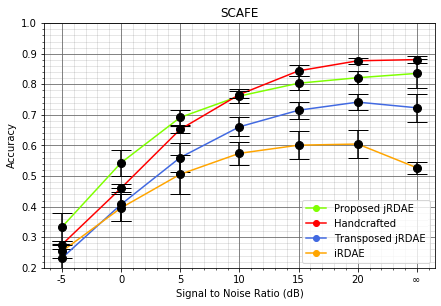}
		\label{label-4}
	\end{minipage}
	\begin{minipage}[b]{.48\textwidth}
		\vspace{0pt}
		\includegraphics[width=\columnwidth]
		{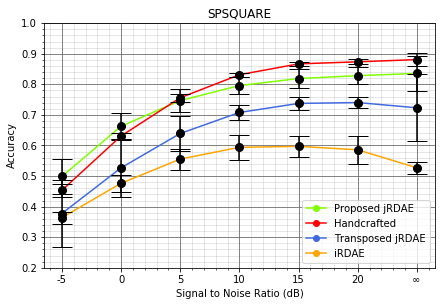}
		\label{label-5}
	\end{minipage}
	\begin{minipage}[b]{.48\textwidth}
		\vspace{0pt}
		\includegraphics[width=\columnwidth]
		{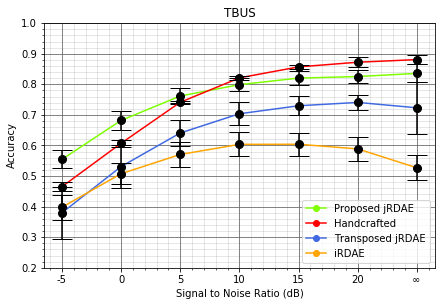}
		\label{label-6}
	\end{minipage}
	\caption{Accuracy results itemized by additive noise and SNR for different architectures. Confidence intervals are also depicted for each of the results taken as one standard deviation on the 3-fold validation.}
	\label{fig:results}
	\vspace{-1.5em}
\end{figure*}

\fi

Our results are displayed in Figure \ref{fig:results}. As a metric to compare the algorithms, we chose Accuracy in terms of speaker identification as the classes were fairly balanced. \textcolor{black}{For each experiment, the confidence interval is shown as a small box-and-whisker plot representing the standard deviation of the cross-validation experiments performed to indicate its statistical significance.} Our aim is to achieve robustness and therefore to obtain a less degraded performance when the SNR is low. 

The independently optimized cascaded architecture (iRDAE) is the algorithm that achieves the lowest results at all SNRs (with the exception of OHALLWAY at SNRs lower than 10 dB where it is the second worst). We can conclude that the optimization of the RDAE only, towards minimizing MSE is not consistent with the needs of the SI. 

The transposed architecture is the result of taking the spectrograms' axis transposed and therefore reducing the time axis in the autoencoder instead of the frequency. As can be seen in the plots, this results in an inaccurate detection of the speaker. We believe that reducing the sequential temporal characteristics of the spectrograms is a handicap for the SI system.

The handcrafted-features (HC), on the other hand, achieve good results for high SNRs, since the features where chosen specifically for the task. HC works acceptably well when small amount of data is available, but its performance worsens very fast when SNR decreases.

For most of the noises, the proposed architecture (jRDAE) achieves the best results for lower SNRs and stable rates for higher ones. jRDAE achieves reliable results for the whole range of SNRs, being a more robust approach than the rest of the architectures. The exception is the PRESTO noise in which a closer look revealed that the denoised spectrograms where rather far from the clean ones. 

Additionally, we stratify the results for the proposed jRDAE system (Table \ref{table:results_stress})
to observe the differences in performance for \textit{neutral} (N) and \textit{stressed} (S) samples. Clearly, lower SI rates were observed in stressed utterances, showing the difficulties induced by stress, PRESTO and SCAFE being the most affected. This suggests the need to specifically cater for distortions caused by emotional speech as we outline in \ref{sec:Conclusions}. 

\ifdefined\imageson
\begin{table}[h]
	\vspace{-2em}
	\hspace{-1.7em}
	\footnotesize
	\begin{tabular}{|c|c|ccccccc|c|c|}
		\hline
		\multicolumn{2}{|c|}{Noise \textbackslash~ SNR} & \multicolumn{1}{c|}{-5} & \multicolumn{1}{c|}{0} & \multicolumn{1}{c|}{5} & \multicolumn{1}{c|}{10} & \multicolumn{1}{c|}{15} & \multicolumn{1}{c|}{20} & Clean   & \textbf{Mean} & \textbf{Std} \\ \hline
		\multirow{2}{*}{DWASHING}          & N         & 36.60                   & 56.04                  & 69.23                  & 78.37                   & 81.77                   & 83.78                   & -     & 67.63         & 1.98         \\ \cline{2-2}
		& S         & 28.45                   & 45.58                  & 58.54                  & 68.88                   & 74.71                   & 78.47                   & -     & 59.11         & 1.14         \\ \hline
		\multirow{2}{*}{OHALLWAY}          & N         & 49.00                   & 68.76                  & 78.09                  & 81.96                   & 83.87                   & 85.27                   & -     & 74.49         & 2.42         \\ \cline{2-2}
		& S         & 43.43                   & 60.98                  & 71.17                  & 76.74                   & 79.98                   & 81.44                   & -     & 68.96         & 1.28         \\ \hline
		\multirow{2}{*}{PRESTO}            & N         & 28.53                   & 45.92                  & 65.59                  & 73.85                   & 79.58                   & 82.94                   & -     & 62.74         & 1.91         \\ \cline{2-2}
		& S         & 20.33                   & 38.14                  & 56.84                  & 68.60                   & 75.01                   & 78.63                   & -     & 56.26         & 1.02         \\ \hline
		\multirow{2}{*}{TBUS}              & N         & 60.05                   & 72.40                  & 80.14                  & 83.40                   & 85.97                   & 85.87                   & -     & 77.97         & 2.43         \\ \cline{2-2}
		& S         & 53.46                   & 66.37                  & 74.47                  & 78.39                   & 80.34                   & 81.12                   & -     & 72.36          & 1.07         \\ \hline
		\multirow{2}{*}{SCAFE}             & N         & 41.21                   & 61.49                  & 75.29                  & 80.89                   & 84.20                   & 85.59                   & -     & 71.45         & 2.05         \\ \cline{2-2}
		& S         & 29.90                   & 51.25                  & 66.55                  & 74.13                   & 78.71                   & 80.68                   & -     & 63.54         & 1.47         \\ \hline
		\multirow{2}{*}{SPSQUARE}          & N         & 54.08                   & 71.42                  & 78.97                  & 83.03                   & 85.22                   & 85.45                   & -     & 76.36         & 2.9         \\ \cline{2-2}
		& S         & 48.05                   & 64.05                  & 72.82                  & 78.11                   & 80.46                   & 81.70                   & -     & 70.87         & 1.58         \\ \hline
		\multirow{2}{*}{CLEAN}             & N         & -                       & -                      & -                      & -                       & -                       & -                       & 86.29 & -             & -            \\ \cline{2-2}
		& S         & -                       & -                      & -                      & -                       & -                       & -                       & 82.41 & -             & -            \\ \hline
	\end{tabular}
	\caption{Accuracy results for stratification of Stressed (S) and Neutral (N) samples on Speaker Identification for proposed jRDAE. \textcolor{black}{In the last two columns, \textit{mean} and \textit{std} values are provided as a summary. The \textit{std} values denote the average of the \textit{std} values of the 3-fold validation process for the 6 SNRs.}}
	\label{table:results_stress}
\end{table}
\fi

\ifdefined\imageson
\begin{figure*}[!h]
	\centering
	\begin{minipage}[b]{.48\textwidth}
		\vspace{0pt} 
		\includegraphics[width=\columnwidth]
		{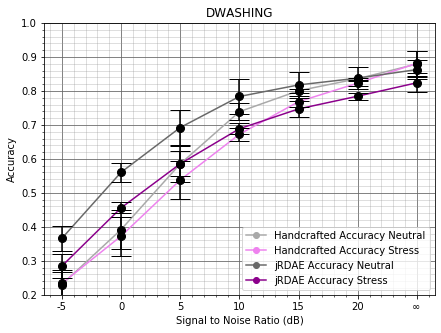}
		\label{label-7}
	\end{minipage}
	\begin{minipage}[b]{.48\textwidth}
		\vspace{0pt} 
		\includegraphics[width=\columnwidth]
		{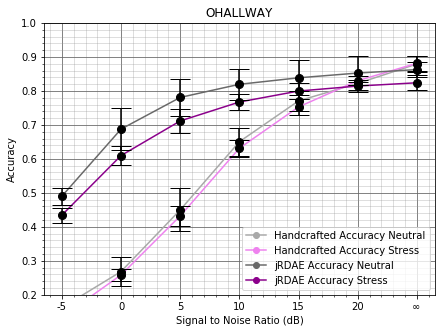}
		\label{label-8}
	\end{minipage}
	\begin{minipage}[b]{.48\textwidth}
		\vspace{0pt}
		\includegraphics[width=\columnwidth]
		{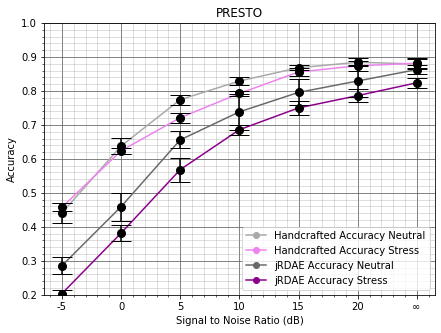}
		\label{label-9}
	\end{minipage}
	\begin{minipage}[b]{.48\textwidth}
		\vspace{0pt}
		\includegraphics[width=\columnwidth]
		{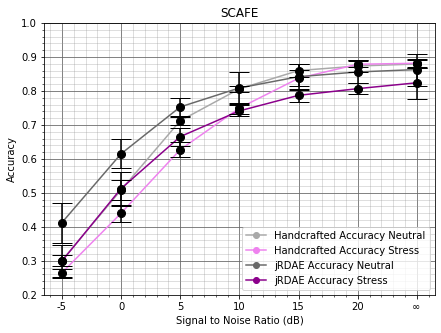}
		\label{label-10}
	\end{minipage}
	\begin{minipage}[b]{.48\textwidth}
		\vspace{0pt}
		\includegraphics[width=\columnwidth]
		{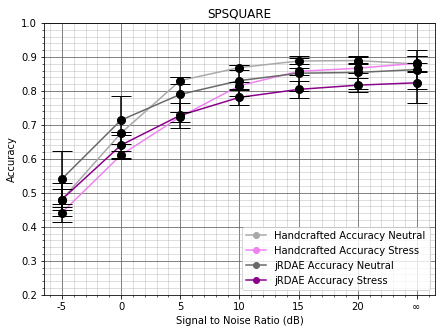}
		\label{label-11}
	\end{minipage}
	\begin{minipage}[b]{.48\textwidth}
		\vspace{0pt}
		\includegraphics[width=\columnwidth]
		{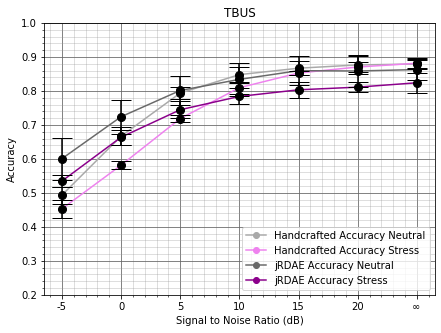}
		\label{label-12}
	\end{minipage}
	\caption{\textcolor{black}{Accuracy results itemized by additive noise and SNR for stress and neutral samples for Handcrafted and jRDAE configurations. Confidence intervals are also depicted for each of the results taken as one standard deviation on the 3-fold validation.}}
	\label{fig:results_stress}
	\vspace{-1.5em}
\end{figure*}
\fi

\textcolor{black}{For the rest of the architectures, we observe a similar deterioration in the stress cases. Specifically for the Handcrafted model, that follows a similar trend as the results in Table \ref{table:results_stress}. Stress accuracy results are slightly worse than neutral ones, and for lower SNRs, the results are notably worse than for jRDAE.}
\textcolor{black}{In Figure \ref{fig:results_stress} we show a breakdown of the results in terms of neutral and stressed speech, comparing the Handcrafted approach and the proposed jRDAE. With a few non significative exceptions, we observe  better results for neutral speech while for stressed speech the SR achieves lower accuracy rates, for both approaches -HC and jRDAE-. This highlights that stress affects speech and deteriorates speaker recognition rates in spite of having included this particular degradation within the training set. In this paper, we are not using the stress vs. neutral labels to actively working to combat stress or reduce its degrading effects and therefore we believe there is room to improve. 
In particular, we plan on developing adversarial strategies for this purpose.}

\textcolor{black}{On the other hand and regarding the system's computational cost, we need to take into account that this speaker identification module is expected to be embedded into a cyber-physical device that is computationally constrained due to the need to lengthen the life of the battery. The decision of using GRU cells instead of LSTM cells was based on the fact that the number of parameters is significantly smaller and therefore GRUs are fast and computationally less expensive than LSTM. With this decision the main speed bottleneck is now the SNN, with 1.1 million parameters. In the future, we aim to reduce the number of parameters of this model to develop a lightweight intelligent algorithm optimizing performance.}



\section{Conclusions \& Future Work}
\label{sec:Conclusions}
In this paper we evaluated the performance of speaker oriented embeddings extracted with an end-to-end architecture composed of a Recurrent Denoising Autoencoder and a Shallow Neural Network. This representation learning based method takes advantage of the joint optimization of both blocks with a combined loss function for the RDAE that incorporates the speaker cross-entropy loss to the MSE employed for denoising. 
This is shown to work better than a general purpose denoiser.  

To further analyse the robustness of this speaker oriented embeddings and end-to-end architecture we aim to test it in an adversarial fashion by using an emotion --or stress-- classifier as a domain adversarial module. We also intend to use richer datasets that contain real life speech, specifically emotions such as panic and fear. In this sense, UC3M4Safety Group is currently collecting a dataset which records real stress and fear emotions induced in women. 
To deal with the problem of data scarcity we plan to transfer models and adapt to emotional speech other large-scale datasets for speaker identification such as the crowd-annotated VESUS \citet{VESUS} and VOXCeleb \citet{voxceleb}.







\begin{acknowledgements}
The authors would like to thank the rest of the members of the UC3M4Safety for their support and NVIDIA Corporation for the donation of a TITAN Xp. 
This work has been partially supported by the Dept. of Research and
Innovation of Madrid Regional Authority (EMPATIA-CM  Y2018/TCS-5046) and 
the Dept. of Education and Research of Madrid Regional Authority with a European Social Fund for the Pre-doctoral Research Staff grant for Research Activities, within the CAM Youth Employment Programme (PEJD-2019-PRE/TIC-16295). 

This is a preprint of an article published in Neural Computing and Applications, Springer. The final authenticated version is available online at: https://doi.org/10.1007/s00521-021-06083-7
\end{acknowledgements}

\noindent{\footnotesize\textbf{Conflict of interest} The authors have no conflicts of interest to declare that are relevant to the content of this article.}

%
%

\bibliographystyle{spbasic}      
\bibliography{references}   

%
%

\end{document}